\documentstyle[12pt]{article}
\begin{document}
\baselineskip 18pt
\newcommand{\Dirac}{/\!\!\!\!D}
\newcommand{\beq}{\begin{equation}}
\newcommand{\eeq}[1]{\label{#1}\end{equation}}
\newcommand{\bea}{\begin{eqnarray}}
\newcommand{\eea}[1]{\label{#1}\end{eqnarray}}
\renewcommand{\Re}{\mbox{Re}\,}
\renewcommand{\Im}{\mbox{Im}\,}
\begin{titlepage}
\hfill  hep-th/9611201, NYU-TH-96/11/01, IASSNS-HEP-96-117
\begin{center}
\hfill
\vskip .4in
{\large\bf M-Theory Origin of Mirror Symmetry in Three Dimensional
Gauge Theories}
\end{center}
\vskip .4in
\begin{center}
{\large Massimo Porrati}
\vskip .1in
{\em Department of Physics, NYU, 4 Washington Pl.,
New York, NY 10003, USA\\ and\\ Rockefeller University, New York, NY
10021-6399, USA\footnotemark}
\footnotetext{e-mail porrati@mafalda.physics.nyu.edu}
\vskip .1in
and
\vskip .1in
{\large Alberto Zaffaroni}
\vskip .1in
{\em Institute for Advanced Study, Olden Lane, Princeton, NJ 08540,
USA\footnotemark}
\footnotetext{e-mail zaffaron@sns.ias.edu}
\end{center}
\vskip .4in
\begin{center} {\bf ABSTRACT} \end{center}
\begin{quotation}
\noindent
We present M-theory compactifications on $K_3 \times K_3$ with
membranes near the $A_n$ or $D_n$ singularities of the $K_3$ spaces.
By realizing each of these compactifications in
two different ways as type I' models with 2- and 6-branes,
we explain the three-dimensional duality between
gauge theories recently found by Intriligator and Seiberg. We also find
new pairs of dual gauge theories, which we briefly describe.
\end{quotation}
\vfill
\end{titlepage}
\eject
\noindent
\section{Introduction}
A recent study of three dimensional gauge theories with $N=4$ supersymmetry,
from the viewpoint of both field~\cite{WS,H} and string~\cite{IR} theory,
led to the discovery~\cite{SI} of a new duality between two classes of gauge
theories. The first one is a $U(1)$ gauge theory with $n+1$ electrons, or
$SU(2)$ with $n$ quarks. The global symmetry is associated to the
ADE classification: $SU(n+1) = A_n$ and
$SO(2n)=D_n$, respectively. The second one is the gauge theory associated
with Kronheimer's hyper-K\"{a}hler quotient construction of ALE
spaces~\cite{K}. The duality exchanges theories corresponding to the same ADE
group-singularity.

N=4 theories are the dimensional reduction of N=1 theories in six
dimensions. Their global R-symmetry,
$SO(4)\cong
SU(2)_L\times SU(2)_R$, can be related to the $SU(2)_R$ six-dimensional
R-symmetry, and to  the $SU(2)_L$
rotation of the three scalars obtained in the dimensional
reduction of the vector. Both the Higgs and the
Coulomb branch are hyper-K\"{a}hler
manifolds, and the Higgs branch is not renormalized by quantum effects. Two
kinds of coupling constants can be added: masses (vector of $SU(2)_L$,
they lift part of the Higgs branch), and Fayet-Iliopoulos terms (vector for
$SU(2)_R$, they lift part of the Coulomb branch).

The duality discovered in~\cite{SI}, as usual in non-finite theories,
relates long distance effective theories.
It simultaneously exchanges the Coulomb branch with the Higgs branch,
the mass terms with the Fayet-Iliopoulos terms, and $SU(2)_R$ with $SU(2)_L$.

In this paper we want to relate this picture to a string theory
compactification
in which this duality is reintepreted as a string-string duality, and the
parameters are identified with the VEVs of spacetime background fields.
We will also gain a deeper understanding of this duality by
considering a compactification of M-theory on $K_3\times K_3$, with all
membranes near an $A_k \times A_n $ or a $D_n \times D_k$ singularity of
$K_3\times K_3$. In
M-theory, the three-dimensional duality follows from realizing in two
different ways the same M-theory configuration in terms of type I' models
with 2- and 6-branes.
Besides its simplicity, this M-theory explanation can be implemented in
details in terms of a fairly simple
construction involving 6- and 2- D-branes in
type IIA theory on $T^3/\Omega Z_2 \times R^4/Z_n$.
This must be contrasted
with another space-time explanation of the duality, which uses T-duality
between type IIA and IIB string theory. In the latter approach, the
identification of the Calabi-Yau spaces giving rise to the three-dimensional
gauge theories is far from trivial.

 Our M-theory explanation of the dualities in ref.~\cite{SI} can be
easily generalized to obtain a larger class of $N=4$ gauge-theory dual
pairs, which are related to the quiver construction
in ref.~\cite{DM}.\footnotemark
\footnotetext{As we were completing this work,  a paper~\cite{oz} appeared
which covers related topics. In particular, \cite{oz}  identifies the
Calabi-Yau pairs that explain some of the dualities of ref.~\cite{SI}
via type IIA-IIB  T-duality, and construct
new examples of mirror pairs
starting from quivers diagrams. As we will comment below, some
of these new examples can be naturally understood in the context of our
construction.}

Both theories in ~\cite{SI} can be interpreted as the world-volume
theory for a probe~\cite{D,S,DSS} in a string spacetime context.
A more general construction~\cite{DM}, associated to quiver diagrams,
contains both of them as subsectors, and, as we will see, is the natural
setting for explaining the duality and for finding new dual pairs.
For simplicity, we will begin in section 2 and 3 by
reviewing the construction of the theories in~\cite{SI}. After having
understood these
building-blocks, it will be easy to construct the general quiver theory
in section 4 and to show how to get the result of ref.~\cite{SI}. In section 5
we will generalize this construction to find new dual pairs.
\section{The Type I' Probe}
Let us start with what we will call in the following the type I'-probe theory.
Following~\cite{IR}, we will
consider type IIA on $T^3/\Omega Z_2$, where $Z_2$
changes the sign to all coordinates in $T^3/\Omega Z_2$.
This orientifold is better
known as type I' theory, the T-dual of type I. There are 16 D-6-branes
(plus 16 images under $\Omega Z_2$)
which fill the non-compact space.\footnotemark
\footnotetext{Their position in $T^3/\Omega Z_2$ corresponds to the
Wilson lines of $SO(32)$ in the original type I picture.}
 Type I' on $T^3/\Omega Z_2$
is conjectured to be dual to M-theory compactified on
$K_3$. The gauge group is
generically $U(1)^{16}$ but it is enhanced to $U(n)$ when the positions of
$n$ 6-branes coincide, and to $SO(2n)$ when $n$ 6-branes and their $n$ images
under $\Omega Z_2$
coincide at an orientifold point. At the corresponding point in moduli
space, M-theory is conjectured
to develop an $A_{n-1}$ and a $D_n$ singularity, respectively. Consider now
a 2-brane probe transverse to the compact space.\footnotemark
There is generically
a $U(1)$ gauge theory on the world-volume, which is enhanced to $SU(2)$
when the probe meets its image at an orientifold point.
\footnotetext{In M-theory this is a membrane transverse to $K_3$, while in
the T-dual type I picture
it is the 5-brane corresponding to a small instanton wrapped
around $T^3$.}
The three transverse scalars in $T^3/\Omega Z_2$
are the partners of the gauge field, the
other four form a hypermultiplet, X, which is decoupled~\cite{Ws,GP}. The open
strings which connect 2- and 6-branes are
hypermultiplets $\phi_i$, $i=1,..n$,
on the world-volume of the probe and have a
mass proportional to
the distance between the branes. It is now easy to realize
a combined spacetime-probe situation in which the probe world-volume
describes
exactly the three dimensional theory we are looking for~\cite{S}.
In the region of
moduli space near a $U(n)$ enhanced spacetime gauge symmetry, a probe in the
region of the $n$
spacetime branes describes a world-volume gauge theory of $U(1)$ with $n$
electrons,
while near an $SO(2n)$ spacetime gauge symmetry, the world-volume theory is
$SU(2)$ with $n$ quarks. The spacetime gauge symmetry becomes the global
flavor symmetry on the world-volume.

Let us consider in some detail the case
of
$U(1)$ with $n$ electrons.
The masses of the electrons are the positions $m_i$ of
the $n$ 6-branes in $T^3/\Omega Z_2$,
one of them can be eliminated by shifting the origin
of the Coulomb branch. When all masses are different,
the Higgs branch of the theory is completely lifted. Let us
denote with $\vec x$ the position of the probe in
$T^3/\Omega Z_2$.
We will work with a very large $T^3/\Omega Z_2$
and well far away from the orientifold points.
The metric in the Coulomb branch is corrected by quantum
effects to~\cite{WS,SI}:
\beq
ds^2=g^{2}(\vec x)(dt+\vec\omega\cdot d\vec
x)^2+g(\vec x)^{-2}d\vec x\cdot d\vec x,
\eeq{1}
with
\beq
g^{-2}(\vec x)={g_{cl}}^{-2}
+\sum _{i=0}^{n-1}{1\over |\vec x-\vec m_i|},\qquad \vec\nabla
(g^{-2})=\vec\nabla   \times \vec \omega.
\eeq{2}
The classical Coulomb branch is
$
R^3\times S^1$ where the radius of $S^1$ is given by $g_{cl}$. The metric is
1-loop exact~\cite{WS}, and changes
the topology at infinity
to $R^4/Z_n$, thanks to $\vec\omega$. The metric in eq.~(\ref{1})
is identified locally with the metric on $K_3$ in the
M-theory description; the 2-brane probe is identified with a membrane, whose
world-volume theory
indeed describes a sigma model on $K_3$ with metric given by eq.~(\ref{1}).
For generic masses, the 6-branes are separated and this metric
(as well as
the corresponding $K_3$) is smooth. When masses coincide, there is
an $A_{n-1}$ singularity with non-trivial metric~\cite{WS,a}; only when
$ g_{cl}$ goes to infinity, does this correspond to a flat $R^4/Z_n$ space.
Note
that the limit $ g_{cl}\rightarrow\infty$, for generic $\vec m_i$, gives
an ALE space (see eq.~(\ref{1'}) below).

The metric in eq.~(\ref{1}) has been obtained with a world-volume
computation,
but it could have been obtained as well using spacetime considerations. The
2-brane probes the value of the spacetime dilaton at the point
$\vec x$~\cite{DSS,S2}.
The spacetime 6-branes provide delta function sources for the equation of
motion of the dilaton, which now depends on the three coordinates of
$T^3/\Omega Z_2$;
the solution of the
Laplace equation in three dimensions with sources in $\vec m_i$
gives to the dilaton a dependence on the inverse of the distance exactly as
in eq.~(\ref{2}). The coupling constant $g_{cl}$ corresponds
to the asymptotic value of the dilaton.

When the masses coincide (and thus
can be put to zero by shifting the origin of
$\vec x$), the Coulomb branch has a singularity from which a
$n-1$-dimensional (in quaternionic unity) Higgs branch starts. This has been
interpreted~\cite{Ws,D} as the moduli space of $U(n)$ instantons on $R^4$.
In fact,
the Coulomb branch describes the phase in which the 2-brane is separated from
the background branes. To go to the Higgs branch we need to tune $\vec x$ and
$\vec m_i$ to zero; this implies that the background branes coincide
--thereby giving
a $U(n)$ spacetime gauge group-- and that the probe meets the 6-branes.
In this configuration, the probe is located at a point
on the (coinciding) background branes, and represents
a zero-size instanton of their
$U(n)$ world-volume gauge group. The gauge field coupling
to the RR background three form $A^3$ reads $F\wedge F\wedge A^3$. This
shows indeed
that we can trade a source for $A^3$ (the 2-brane) for an instanton~\cite{D}.
We can now give expectation value to the hypermultiplets $\phi_i$; this
corresponds to give a finite, nonzero size to the instanton.
This mechanism applies in general to $p$ and $p+4$ parallel branes. The Higgs
branch is not corrected by quantum effects and can be determined classically
by solving the algebraic equations for the D-terms.

The $D_n$ case is quite similar. Spacetime branes and probe meet now near an
orientifold point.
The Coulomb branch describes a space which at infinity has the
topology of $R^4/D_n$, and is smooth for generic masses. When the spacetime
branes coincide, there is a singularity from which a one-dimensional
Higgs branch starts,
corresponding to the moduli space of $SO(2n)$ instantons. In the
type I description this is exactly the small instanton described in~\cite{Ws}.
\section{The Kronheimer Construction}
Consider now the gauge theory corresponding to the Kronheimer construction.
Using the same notations as in~\cite{SI}, the gauge group is $K_G = (\prod
_{i=0}^rU(n_i))/U(1)$,
where $i$ runs over the nodes of the extended Dynkin diagram of the
group $G$ of rank $r$, and $n_i$ is the Dynkin index of the node. The
diagonal $U(1)$ is not gauged. The
hypermultiplet  content is $\oplus _{ij}a_{ij}(n_i,n_j)$, where
$a_{ij}$ is the adjacency matrix for the extended Dynkin diagram.\par
If $G$ is simply laced, this gives a construction of the ALE spaces. There is
a one-dimensional Higgs branch. The solution of the D-term equations gives the
singular space $R^4/\Gamma_G$, where $\Gamma_G$ is the discrete group
associated to $G$. We can blow up the singularity by
introducing Fayet-Iliopoulos terms $\vec \zeta$ in the gauge theory,
one for each $U(1)$, with the constraint that their sum is zero.
The solution of the D-term equations determines the metric on the resolved
ALE space, which depends on the
parameters $\vec \zeta$. For instance,
in the natural variables which solve the
D-term equations, the metric for the $Z_n$ case is given by:
\beq
ds^2=g^{2}(\vec x)(dt+\vec\omega\cdot d\vec
x)^2+g(\vec x)^{-2}d\vec x\cdot d\vec x,
\eeq{1'}
with
\beq
g^{-2}(\vec x)=
\sum _{i=0}^{n-1}{1\over |\vec x-\vec m_i|},\qquad \vec\nabla
(g^{-2})=\vec\nabla   \times \vec \omega.
\eeq{2'}
\par
These $N=4$ gauge theories can be constructed in terms of
D-branes~\cite{DM,P,J}.
Consider type IIA ``compactified'' on the space $R^4/\Gamma_G$ and put a
certain number of 2-brane probes transverse to the singular space. We
need  $g=|\Gamma_G|$ branes in order to have a faithful representation of the
discrete group. If we project out the Chan-Paton factor of the gauge fields
and of the hypermultiplets --corresponding to the positions of the probes in
$R^4/\Gamma_G$-- with the
$g$-dimentional regular representation of the orbifold
group, we obtain exactly the spectrum of the $K_G$ gauge theory.
In the $Z_n$ case, that we shall study next,
the regular representation (which is always block
diagonal, and contains
every representation a number of times equal to its
dimension) is $\gamma(\zeta ) = diag\{1,\zeta ,...,\zeta^{n-1}\}$ where
$\zeta=\exp(2\pi i/n)$.
The bosonic fields on the world-volume are the gauge fields
$A_{\mu}$, together
with their scalar partners (the transverse position $\vec x$ in $T^3/\Omega
Z_2$),
and the hypermultiplets X, describing positions in the ALE space. We will write
them as the two complex scalars, $X^1$, $X^2$, which diagonalize the action of
$\Gamma_G$.
The $n\times n$ Chan-Paton factors of the gauge fields
and the two complex scalars must satisfy:
\bea
\gamma\lambda_{A_{\mu}}\gamma^{-1} &=& \lambda_{A_{\mu}}\nonumber\\
\gamma\lambda_{X^1}\gamma^{-1} &=& \zeta\lambda_{X^1}\nonumber\\
\gamma\lambda_{X^2}\gamma^{-1} &=& \zeta^{-1}\lambda_{X^2},
\eea{3}
giving non-zero entries only for $A_{ii}$, together with its three scalars
partners $\vec{x}_i$, and $X^1_{i-1,i}$, $X^1_{n-1,1}$, $X^2_{i,i-1}$,
$X^2_{1,n-1}$. These fields describe an N=4 gauge multiplet
$U(1)^n$, and hypermultiplets charged only under adjacent $U(1)$s
(the first and
the last $U(1)$ are considered adjacent, in the spirit of the extended Dynkin
diagram construction). This is exactly the spectrum of $K_{SU(n)}$.

The singular space can be blown up by turning on VEVs of
twisted fields. There are $n-1$
twisted vector fields $A_{\mu}^{Bi}, i=1,...,n-1$, coming from the orbifold
point. It can be easily seen from the conformal field theory on the orbifold
that these are RR fields.\footnotemark
\footnotetext{A shotcut is to consider that the RR fields come, in the
compactification on the resolved space, by reducing the RR 3-form of
type IIA on the $n-1$ spheres that resolve the singularity.}
Since they are RR background fields, they couple to the
world-volume gauge fields as~\cite{D,DM}:
\beq
\int d^3x \sum_{i=1}^{n-1} A^{Bi}\wedge tr(\gamma (\zeta^i)F).
\eeq{c}
The supersymmetrization of this coupling implies that the three spacetime
partners of $A^{Bi}_\mu$ (responsible for the blow-up of the orbifold)
behave as FI terms for the branes~\cite{DM}. In this way we get a natural
correspondence between
the spacetime orbifold construction and the Kronheimer gauge theory. The motion
of a probe on the resolved space should describe a sigma model on the ALE.
The Higgs branch of $K_G$,
in which only one hypermultiplet is left massless,
 provides such a description.
\section{The M-Theory Interpretation}
At this point, it is not surprising, maybe, that the
two gauge theories~\cite{SI} are related by duality.
Start with type I' on $T^3/\Omega Z_2$, with $n$ near-coincident
spacetime branes at a generic point, or $2n$
near-coincident branes at an orientifold point.
String duality relates this compactification to M-theory compactified
on a $K_3$ which is locally an ALE space
(associated to $A_n$ or $D_n$)~\cite{IR}. The probe is
now
a membrane of M-theory. Compactifying  on a circle (in a direction
transverse to the brane) we get exactly type IIA on an ALE space with a 2-brane
probe. The two gauge theories are mapped into each other by the string
duality.

A deeper description, more detailed and symmetric, is obtained by working
always in type I' theory. This description will allow us to find the M-theory
origin of the three dimensional mirror-symmetry.
To this purpose, we must analyze a more general construction.

Consider the theory of the probe in type I' on $T^3/\Omega Z_2$
and ``compactify'' the
remaining four transverse directions on an ALE space. This model has been
considered in~\cite{DM} and, after a T-duality, corresponds exactly to the
construction in~\cite{GP}, where now the singular $K_3$ is replaced by
a singular ALE. The non-compactness of the space still allows us to consider
the 2-brane as a probe; this implies, for example,
that we do not have to impose
the tadpole conditions (the 2-brane flux can escape to infinity in the ALE),
so that the number of probes and their world-volume gauge
group are essentially unconstrained.
Using the same notation as in the two examples above, which can be though of as
two limits of the present construction, we now need $2g$
branes to account for the images under $\Omega Z_2$ and
$\Gamma_G$. The hypermultiplets X, which were decoupled in the type I'
example, now become a coupled set of hypermultiplets,
describing the position of the system on the ALE space, as in the
$K_G$ example. We still have
hypermultiplets, $\phi$, which correspond to open string connecting 2- and
6-branes. This
set of fields now couples in a complicated way. To work out the field content
of the world-volume gauge theory, we must first find out the way in which
 $\Omega Z_2$ and
$\Gamma_G$ act on the Chan-Paton factors. This action is constrained by some
consistency conditions listed in~\cite{GP}. The solution of these conditions
and the corresponding world-volume theory can be found in~\cite{DM},
for several
particular cases where it is related to a quiver diagram construction.

Consider, to begin with,
the case of $k$ near-coincident 6-branes at a generic point on $T^3/\Omega
Z_2$,
and a compactification on an ALE with an $A_n$ singularity.
This means that there is
a $U(k)$ spacetime gauge symmetry, and that we do not need to project the
probe theory with $\Omega Z_2$. The Higgs branch of this theory has been
interpreted~\cite{DM} as the moduli space of a $U(k)$ instanton
on the ALE space. This follows from the general construction we sketched in the
discussion of the type I' example. In~\cite{DM}, it
was shown that the solution
of the D-term equations indeed corresponds to a mathematical construction of
that moduli space, introduced by Kronheimer and Nakajima~\cite{KN}.

Let us work out the field content for the probe world-volume theory. Consider
$n$ 6-branes and a compactification on the ALE space corresponding to
the discrete group $Z_n$. $\Omega Z_2$ relates branes which are far away and
that contribute only extremely massive fields; since it relates 2-6 open
strings with 6-2 strings it does not affect the hypermultiplets $\phi$. We
only have to project out $n\times n$ Chan-Paton factors by $Z_n$, whose action
as an $n\times n$ unitary matrix has been given in the discussion of the
$K_G$ theory. The projection for the gauge fields and the hypermultiplets
$X$
is as in eq.~(\ref{3}) and gives the same result: $U(1)^n$ with
hypermultiplets charged under adjacent $U(1)$s.
The hypermultiplets $\phi_{i,A}$,
which have an index $i$ in the fundamental of the global spacetime symmetry
$U(k)$, and a Chan-Paton index $A$, are projected by
$\phi_{iA}=\gamma(\zeta )_{AB}\phi_{iB}$, giving $k$ extra hypermultiplets
charged only with respect to the first $U(1)$.\par
The VEVs of spacetime fields become world-volume coupling constants.
The classical coupling $g_{cl}$ is the asymptotic value of
the dilaton; the masses for the $\phi_i$ are the positions of the spacetime
6-branes (there are only $k-1$ independent mass vectors since we have the
freedom to shift the origin of the Coulomb branch); the FI terms for
the $X$ are the $n-1$ twisted-sector blow-up vector moduli of the ALE.
The global R-symmetry
group of the N=4 theory is easily identified, by relating it to a
dimensional reduction from 10 dimensions. Namely,
$SU(2)_L$  corresponds to the rotations
on $T^3/\Omega Z_2$,
and $SU(2)_R$ corresponds to the subgroup of $SO(4)$
(rotation of the coordinates
corresponding to $X$) unbroken by the ALE. It is clear from this spacetime
interpretation that masses are vector of $SU(2)_L$ and FI are vectors of
$SU(2)_R$.

Let us write the D-term equations for the Higgs phase of this theory. Since
we have implicitly used a trivial action of $Z_n$ on the background branes
(which are ``wrapped'' on the ALE space) we are assuming, according
to~\cite{DM}, that we are dealing with the moduli space of $U(n)$ instantons on
the ALE that do not break the gauge group at infinity. The D-term equations
read
($\phi$ is
decomposed in two complex scalars $\phi^1,\phi^2$):
\bea
\sum_{j=0}^{k-1}(|\phi^1_j|^2 -|\phi^2_j|^2) + |X^1_{0,1}|^2 -|X^1_{n-1,0}|^2 +
|X^2_{0,n-1}|^2
- |X^2_{1,0}|^2 &=& \zeta_1^D\nonumber\\
 |X^1_{i,i+1}|^2 -|X^1_{i-1,i}|^2 + |X^2_{i,i-1}|^2
- |X^2_{i+1,i}|^2 &=& \zeta_i^D,\qquad i=1,...,n-1\nonumber\\
\sum_{j=0}^{k-1}(\phi^1_j\phi^2_j) + X^1_{0,1}X^2_{1,0} -
X^1_{n-1,0}X^2_{0,n-1}
&=&
\zeta_1^F\nonumber\\
X^1_{i,i+1}X^2_{i+1,i} - X^1_{i-1,i}X^2_{i,i-1} &=& \zeta_i^F,\qquad
i=1,...,n-1,\nonumber\\
\eea{Dterm}
where $n \equiv 0$.
The $n-1$ independent FI terms are associated to  a set of
collapsing two-spheres, whose intersection matrix is the Cartan matrix of $G$.
This means that
in this basis $\sum_i \zeta_i=0$. By adding up all these equations,
it is immediately seen that
the contribution of the $\phi$ and that of the $X$ factorize; in particular the
equation for $\phi$ is not affected by the FI terms, when they satisfies the
previous condition. Since we have to mod out by the gauge group, this moduli
space is a fibration over the moduli space of
a $U(k)$ instanton on $R^4$.

The equations determining the vacuum manifold, in a general phase (Higgs or
Coulomb) are eq.~(\ref{Dterm}) and:
\bea
(\vec{x}_0-\vec{m}_j) \cdot\vec{\sigma}_{AB}\phi_j^B&=&0,\qquad
j=0,..,k-1, \qquad A,B=1,2,
\nonumber\\
(\vec{x}_i-\vec{x}_{i+1})\cdot \vec{\sigma}_{AB}X^B_{i,i+1}&=&0,\qquad
i=0,..,n-1.
\eea{Fterm}

Using string duality, we can relate this compactification (in a suitable
region of moduli space) to M-theory on the product of two large $K_3$ spaces,
that, when the probes are near a singularity, can be approximated by two
ALE spaces. The
$k-1$ masses for the 6-branes parametrize the moduli of the first ALE,
while the $n-1$ FI parameterize the moduli of the second.
This M-theory compactification can be described in two different ways in
type I' superstrings; namely, as a compactification on $T^3/\Omega Z_2 \times
R^4/Z_k$ with $n$ 6-branes or as type I' on $T^3/\Omega Z_2\times R^4/Z_n$
with $k$ 6-branes. This two type I' models give rise to two different
three-dimensional world-volume theories on the probe, which, because of this
construction, are obviously dual.
To stay in the region of moduli space
in which this approximation is valid, we must consider a very large
$T^3/\Omega Z_2$ and
we must set $g_{cl}=\infty$ (see remarks after eq.~(\ref{1})), thereby
probing the long distance of the world-volume theory.
As it
should be clear from the spacetime intepretation of the various world-volume
fields, the Coulomb branch describes the motion of the
probe on $T^3/\Omega Z_2$, that is, on the first ALE,
while the Higgs branch is
associated with the motion on the second. As we saw before,
masses and FI are naturally
associated with the first and second ALE, respectively.
In the coordinates used above in eq.~(\ref{1'}), $SU(2)$ acts on the ALE
space by rotating $\vec x$. The $SU(2)_L$
on the probe world-volume
is associated naturally with the $SU(2)$ on the first ALE, while the $SU(2)_R$
is associated with the $SU(2)$ on the second.
Therefore, this
 duality exchanges the Coulomb branch with the Higgs branch, $SU(2)_L$ with
$SU(2)_R$, and the masses with the FI terms.

Let us perform some  simple checks. Turn on both masses and FI terms. This is
a completely geometrical compactification of M-theory; thus, we are free to
exchange the order of compactification. In type I', the fields $\phi$ are
massive (lifting the Higgs branch for the $\phi$)
and the FI terms  lift all the
$U(1)$s except the diagonal one,
under which the $\phi$ are charged. We are left
with the product of the Higgs branch for the theory $K_{SU(n)}$
(depending on the $n-1$
parameters $\vec{\zeta}_i$)
and the Coulomb branch for $U(1)$ with $k$ electrons
(depending on the $k-1$ parameters $\vec{m}_i$).
In the limit in which $g_{cl}$ goes
to infinity, this theory, and the one in which $k$ and $n$ (as well as
$\vec m_i$ and
$\vec \zeta_i$) are interchanged, correspond to the same M-theory construction.
This implies that the Higgs branch for the theory $K_{SU(n)}$
must coincide with
the Coulomb branch for $U(1)$ with $n$ electrons, since
the duality exchanges masses with FI.
This is indeed the simplest
example to check, by simple inspection of eqs.~(\ref{1},\ref{1'}). The check
was done in~\cite{SI}. For large coupling constant
both manifolds are ALE spaces.
This is indeed what we expect from the M-theory picture: the theory on the
membrane is a sigma model on the product of two ALE spaces.

The two theories in~\cite{SI} are contained as subsectors. They can be easily
decoupled by going to infinity on one of the
two ALE (where the
metric is flat). Going to infinity in the first ALE
means sending to infinity the center of mass of the system of branes on
$T^3/\Omega Z_2$, $\vec{x}^{cm}=\vec{M}$,
while keeping finite or zero the relative positions $\vec x_i'$
($\vec x=\vec{M} +\vec x'$).
This decouples completely the fields $\phi$ (their mass is
proportional to $\vec x_1=\vec{M}+...$), but not the fields $X$ (their mass is
proportional to $\vec x_i -\vec x_j$). As we may expect, since the probe is
far from
the background brane, we are left with the $K_G$ theory plus a decoupled
$U(1)$.
With FI terms, the theory is automatically in the Higgs branch (describing
the motion on the
second ALE). By turning off the FI terms, we can go in the Coulomb branch,
which describes, in our picture, a
probe that lies at the singularity of the second ALE and moves
on the first ALE (now locally flat). Going to infinity on the second ALE
means sending to infinity $X$.
We expect on general ground that the D-terms will
factorize in a flat metric for $X$ times the Higgs branch for the type I' probe
theory. In fact, in the region we are considering, the ALE space is now flat
and the instanton moduli space must be the same as the instanton
moduli space on $R^4$. We are left with the type I' probe theory times a flat
sigma model. The Coulomb branch corresponds to the motion of the brane on
the first ALE. When the masses are zero, the theory can be in the Higgs branch,
which means that the probe lies at the singularity on the first ALE, and
moves on the second ALE (now locally flat). This picture is clearly
symmetric under the exchange of the two equivalent realizations of the
M-theory compactification in terms of type I' models. In other words, the
M-theory construction makes  three dimensional gauge theory duality manifest.

The point where the Coulomb and Higgs branches of these theories meet,
which, as argued in~\cite{SI}, gives a non-trivial interacting three
dimensional theory, corresponds
to a point in M-theory where the ALE space is singular,
and where the probe sits on the singularity.
Solitons corresponding to membranes wrapped around vanishing two-spheres
are becoming massless there,
and should be incorporated in the low-energy description.

It would be interesting to understand the spacetime meaning of the remaining
coupling constants: a mass for the Coulomb branch of $K_G$ and a FI term for
Higgs branch of the type I' probe theory. As noted in~\cite{DM}, $\sum_i
\vec{\zeta}_i$
(which we put to zero before) is a natural candidate for the missing FI
term. It would correspond to turning on the VEVs of the
partners of the type IIA 1-form. This form is projected out by the type I'
projection, but if we are far from the
orientifold point the theory is pratically type IIA and this field is
almost massless (this
is consistent with the fact that this FI term can exist only
in the
$A_n$ serie, where the probe is far from the orientifolds).
The D-term equations
still factorize but now this FI enters the equations for $\phi$, deforming
and smoothing the $U(n)$ instanton moduli space, otherwise singular.
The corresponding mass term should be related to the overall mass we always
put to zero by shifting the origin in the Coulomb branch; the precise
implementation of this idea is not yet clear to us.

The $D_n$ series can be discussed along similar lines. Background branes and
probes now lives near an orientifold point.
The probe theory is $SU(2)$ with
$n$ quarks and the $K_G$ is $U(1)^4 \times U(2)^{n-3}$ with hypermultiplets
naturally associated with the extended Dynkin diagram.
The two dual N=4 gauge theories can be obtained in symmetric limits of the
M-theory compactification. Instead of
giving the details, we will briefly explain in the next section how our
construction gives rise to more general examples of three-dimensional
dualities.
\section{New Dual Pairs}

The previous construction can be generalized in the following ways.
First of all,
we can compactify on a generic pair of ALE spaces; by exchanging the two ALEs
we still find a pair of dual gauge theories. Secondly, we
considered complicated gauge theories which reduced, in particular limits,
to the pair of dual
theories in~\cite{SI}. Had we not decoupled them, we would have
found more complex examples of dual pairs. Thirdly, we can consider several
coincident probes. This gives $U(k)$ and $Sp(k)$ groups,
corresponding, in some phases, to
the $k$-instanton sector of the spacetime gauge group on some ALE.
All these models
are related to the quiver diagrams of ref.~\cite{DM}.

Another generalization
may be to relax the condition that
$\Gamma_G$ does not act on the background branes. Since the most general
quiver construction does not have this restriction, it would be interesting
to understand its meaning in terms of our M-theory description.\footnotemark
\footnotetext{ The example called $U(k)^n$ in~\cite{oz} is indeed associated
with
a pair of quiver diagrams of this more general form. Remarkably enough, a
naive application of our method reproduces the dual pairs
explicitly checked in~\cite{oz} . One can speculate that
each quiver diagram can be related to a dual one with a generalization
of the method.}

The generality of this approach should be clear now. Consider a general
``compactification'' of M-theory on two ALE spaces. The interchange of the
two ALEs relates two different gauge theories associated to
different quiver diagrams in type I' theory.
To go to the region of moduli space
in which the M-theory description  is valid, we must  set $g_{cl}=\infty$,
 thereby
probing the long distance of the world-volume theory.
As we stressed before, the Coulomb branch describes the motion of the
probe on the first ALE, while the Higgs branch is
associated with the motion on the second. Masses and FI are naturally
associated with the first and second ALE, respectively;
$SU(2)_L$
is associated naturally with the first ALE, while $SU(2)_R$
is associated with the second.
Therefore, this
 duality exchanges the Coulomb branch with the Higgs branch, $SU(2)_L$ with
$SU(2)_R$, and the masses with the FI.

For instance,
we can see how some of the models in~\cite{oz} fit in our picture.
They are the natural generalization of the probe-$K_G$ theories to the case
of a set of $k$ probes. The dual gauge theories are called A and B, and
described as follows. A model: $U(k)$ gauge group
with $n$ hypermultiplets in the fundamental and one in the adjoint.
B model: $U(k)^n$ gauge group with hypermultiplets in the fundamental of
pairs of adjacent factors, and one hypermultiplet in the fundamental of one
of the $U(k)$. The model B can be
realized in type I' on the $Z_n$ ALE with a background 6-brane. This
corresponds
to M-theory on the ALE space times a smooth $K_3$,
which is locally $R^4$,
(the metric of the Coulomb branch in
type I', which describes a smooth Taub-Nut space, is flat when
$g_{cl}=\infty$). Interchanging the ALEs in the M-theory picture, we get
$k$ probes near $n$ spacetime branes, which give
exactly the content of the A model
(the adjoint hypermultiplet is simply the position of the probes in $R^4$).
Notice that in trying to recover the B model form M-theory directly,
there exists an ambiguity, since superficially we could have put either
one or no spacetime branes in the B model. This ambiguity is resolved by
mapping M-theory on $R^4/Z_n\times R^4$ into the type I' construction
described above.
The $Sp(k)$ case in~\cite{oz} simply corresponds to the $D_n$ serie.

More general dual pairs can be obtained by using two non-trivial ALE spaces,
while keeping the branes at finite distance from the centers of {\em both}
ALEs. We simply quote the simplest examples.

Our main example, $R^4/Z_k\times R^4/Z_n$ (here generalized to the case of
$p$ coincident probes)
leads to a dual pair, in which the first model has a
$U(p)^n$ gauge group with hypermultiplets in the
fundamental of
pairs of adjacent factors, and $k$ hypermultiplets in the fundamental of a
$U(p)$, while the second one has a $U(p)^k$ gauge group
with hypermultiplets in the fundamental of
pairs of adjacent factors, and $n$ hypermultiplets in the fundamental of a
$U(p)$. The dimension of the Higgs and Coulomb branches and the number of
parameters are summarized in Table 2.

$R^4/D_n\times R^4/Z_k$ is complicated because the implementation of
the orientifold projection $\Omega Z_2$ is full of subtleties.
The first model consists of
type I' on $R^4/D_n$ with $k$ background branes; we need not worry about
$\Omega Z_2$, since the 6-branes are far from the orientifolds,
and we easily obtain a theory
$U(1)^4 \times U(2)^{n-3}$ with hypermultiplets
associated with the extended Dynkin diagram and $k$ hypermultiplets
charged under a $U(1)$. The dimension of the Coulomb and Higgs branches are:
$d_C=2n-2$, $d_H=k$. We can put $n+1$ FI terms (one is
associated with an untwisted modulus $\sum\vec\zeta$) and $k-1$ masses.
The dual theory is
type I' with $2n$ spacetime branes near the orientifold ($n$ 6-branes plus
$n$ images),
on $R^4/Z_k$. The consistency condition that $\Omega Z_2$ and
$Z_k$ commute leads to $\gamma_{\zeta}\gamma_{\Omega Z_2}\gamma_{\zeta}^T
= \chi_{\zeta} \gamma_{\Omega Z_2}$, where $\chi_{\zeta}$ is a phase, equal
to 1 for $k$ odd, and to 1 or $\zeta$ (a $k$-th root of unity) if $k$ is even.
The world-volume theories are listed
in full generality in~\cite{DM}. Remarkably, the dimension of the
Coulomb and Higgs branches of all these theories is compatible with duality,
namely: $d_C=k$, $d_H=2n-2$. Let us consider now for simplicity the case $k=2$:
in~\cite{P} a new
consistency condition  for the projections was found, implying that
the only consistent orientifold theory has $\chi = -1$.
  The theory with $\chi = -1$ is
$U(2)$ with 2 hypermultiplets charged under $U(1)$ and $n$ quarks in the
fundamental of $U(2)$. We propose that this theory should be dual to
$U(1)^4 \times U(2)^{n-3}$ with hypermultiplets
associated with the extended Dynkin diagram and 2 hypermultiplets
charged under a $U(1)$. The number of parameters perfectly matches as can be
seen in Table 2. The simplest check of this proposed duality is
to turn on all the parameters associated to the resolution of the two ALE
spaces:
from the point of view of M-theory  we expect that the world-volume theory
is a supersymmetric sigma model on the product of the two
ALEs; this can be indeed checked explicitly in both theories.

For arbitrary $k$ we encounter a subtlety: as noted
in~\cite{P}, the $\Omega Z_2$ operator which is naturally associated with the
orientifold construction~\cite{GJ} is not the limit of the type I'
$\Omega Z_2$ operator of the smooth theory.
The orientifold projection, indeed, intechanges
sectors twisted by $\zeta^i$ with sectors twisted by $\zeta^{k-i}$, projecting
out some of the twisted sectors hypermultiplets; this means that
only some blow-up parameters can be turned on, and that the orbifold
singularity cannot be completely smoothed out.
This is consistent with the fact that in~\cite{DM} it is shown that the D-term
equations are consistent only for $\zeta_{n-i}=\zeta_i$. The consistency
condition in~\cite{P} constrains our choice  of $\chi$ only for $k$ even.
To sum up, the explicit form
of the world-volume theory of $n$ 6-branes near an orientifold of $T^3/Z_2$,
on $R^4/Z_k$, can be found in~\cite{DM} and it is summarized in
Table 1.
Let us simply note that the spacetime
blow-up parameters are in one-to-one correspondence with the FI
terms allowed in the gauge theory. The generic blow-up of the $Z_{k}$
models always leaves $[(k-1)/2]$ $A_1$ singularities; interchanging the
compactification order, this implies that in the dual picture
the spacetime branes must coincide pairwise, and that the world-volume
theory has extra, unbroken global $U(2)$ symmetries.
We conjecture, therefore, that the dual gauge theory is still
$U(1)^4 \times U(2)^{n-3}$ with hypermultiplets
associated with the extended Dynkin diagrams and $k$ hypermultiplets
charged under a $U(1)$, but with only at most the $[k/2]$ independent masses
that respect the global $U(2)$ symmetries.
An alternative possibility is that the dual model is
defined by a $D_n$ projection acting also on the spacetime branes.
The only change in the world-volume
theory is that the $k$ hypermultiplets are now linked to different nodes of
the extended Dynkin diagram, and some of them may belong to doublets of some
$SU(2)$ gauge groups. This reduces the number of allowed masses again to
$[k/2]$~\cite{PZ}.
Whichever the explanation, to obtain the case when masses have generic values
one would have to change in a rather mysterious way --to us, at least--
the definition of $\Omega Z_2$.

The results of our analysis are summarized in Tables 1 and 2.\footnotemark
\footnotetext{
Note that sometimes there is an additional parameter
on the world-volume gauge theory, besides the spacetime blow-up parameters.
This extra parameter may
be associated with the untwisted mode $\sum\vec\zeta$ on the ALE, as we
discussed in
section 4, in which case it can be turned on only when we
are far from the orientifold point.}
Table 1 gives the field content (gauge group and matter representations,
underlined, with multiplicity)
for each of the theories considered in this
paper, while Table 2 gives the dimensions of their Higgs and
Coulomb branches, as well as the number of independent masses and FI
terms. \newpage
\begin{table}[t]\centering
\caption{Field Content of Dual Pairs}
\begin{tabular}{|c|c|c|c|}
\hline Singularity Type & Gauge Group & Matter & Remarks \\ \hline\hline
$Z_k\times Z_n$ & $U(p)^n$ & $ n(\underline{p}^+,\underline{p}^-) + k
\underline{p}^+$ & $p$
2-brane probes,\\ &&&  $\pm$ are $U(1)$ indices \\
\hline\hline $Z_k\times D_n$ & $U(1)^4 \times U(2)^{n-3}$  &
$4(\underline{1}^+,\underline{2}^-) +(n-4)(\underline{2}^+,\underline{2}^-)+
k\underline{1}^+$ & one probe \\ \hline
$D_n\times Z_2$ & $U(2)$ & $2 \underline{1}^+ + n \underline{2}^+$ & \\
\hline
$D_n\times Z_k$ & $SU(2)\times U(2)^{(k-1)/2}$ & $(k-1)/2
(\underline{2}^+,\underline{2}^-)+
1 \underline{1}^+ + n \underline{2}^+$ & $k$ odd \\
\cline{2-4}& $ U(2)^{k/2}$ &
$2 \underline{1}^+ + (k/2-1)(\underline{2}^+,\underline{2}^-) + n
\underline{2}^+$ & $k$ even \\
\hline
\end{tabular}
\vskip 1cm
\caption{Moduli Space and Deformation Parameters of Dual Pairs}
\begin{tabular}{|c|c|c|c|c|c|}
\hline
Singularity Type
& $d_H$ & $d_C$ & $\vec{m}$ & $\vec{\zeta}$ & Remarks \\ \hline\hline
$Z_k\times Z_n$ & $kp$ & $np$ & $k$ & $n$ & \\ \hline\hline
$Z_2 \times D_n$ & 2 & $2n-2$ & 1 & $n+1$ &\\ \hline
$D_n \times Z_2$ & $2n-2$ & 2 & $n+1$ & 1 &\\ \hline \hline
$Z_k \times D_n$ & $k$ & $2n-2$ & $[k/2]$ & $n+1$ & the $k$ 6-branes are\\
&&&&& pairwise coincident \\ \hline
$D_n\times Z_k$ & $2n-2$ & $k$ & $n+1$ & $[k/2]$ & \\ \hline
\end{tabular}
\end{table}

To summarize, in this paper we presented an M-theory construction which
associates pairs of dual three-dimensional gauge theories to quiver
diagrams, and we carried out a preliminary study of some examples.
The case $D_n\times D_k$ can also be studied with the techniques
presented in this paper, and is presently under investigation~\cite{PZ}.
Finally, even though in this paper we considered $R^4/Z_n\times R^4/Z_k$
constructions in which
the spacetime branes are not projected under
$\Gamma_G$, results in~\cite{oz} seems to indicate that even this restriction
could be relaxed~\cite{PZ}.
\vskip .2in
\noindent
{\bf Acknowledgements}\vskip .1in
\noindent
We would like to thank A. Hanany and K. Intriligator for discussions. M.P.
is supported in part by NSF grant no. PHY-9318781; A.Z. is
supported in part by DOE grant no. DE-FG02-90ER40542 and by the
Monell Foundation.

\end{document}